\documentclass[twocolumn,amsmath,amssymb,unsortedaddress,showpacs]{revtex4}
\usepackage{graphicx}
\usepackage{bm}
\newcommand{\mv}[1]{\langle #1\rangle_0}

\begin{document}

\title{Perturbative corrections to the Gutzwiller mean-field solution
of the Mott-Hubbard model}

\author{C. Schroll}
\affiliation{Department of Physics and Astronomy, University of Basel, 
Klingelbergstrasse 82, CH-4056 Basel, Switzerland}

\author{Florian Marquardt}
\affiliation{Department of Physics, Yale University, New Haven, CT 06520, USA}

\author{C. Bruder}
\affiliation{Department of Physics and Astronomy, University of Basel, 
Klingelbergstrasse 82, CH-4056 Basel, Switzerland}

\date{\today}

\begin{abstract}
We study the Mott-insulator transition of bosonic atoms in optical
lattices.  Using perturbation theory, we analyze the deviations from
the mean-field Gutzwiller ansatz, which become appreciable for
intermediate values of the ratio between hopping amplitude and
interaction energy.  We discuss corrections to number fluctuations,
order parameter, and compressibility. In particular, we improve the
description of the short-range correlations in the one-particle
density matrix. These corrections are important for experimentally observed
expansion patterns, both for bulk lattices and in a confining trap
potential.
\end{abstract}


\maketitle

\section{Introduction}
The Mott-Hubbard model of interacting bosons on a lattice has been
used to describe superfluid Mott-insulator transitions in a variety of
systems, e.g., Josephson arrays and granular superconductors
\cite{Fisher01}.  The recent suggestion \cite{Jaksch01} to
experimentally observe this transition in a system of cold bosonic
atoms in an optical lattice and its successful experimental
demonstration \cite{Greiner01} has rekindled the interest in the
Mott-insulator transition and triggered a great deal of theoretical
\cite{Jaksch02,Jaksch03,Hofstetter,Altman,Zwerger01,Buechler01,Damski,Kleinert01,Burnett,Jaksch04,Buechler02,Lewenstein}
and experimental \cite{Greiner02,Mandel01,Esslinger} activity.  The
possibility to directly manipulate and test the many-body behavior of
a system of trapped bosonic atoms in an optical lattice
\cite{Greiner02,Mandel01} is very attractive.  Possible applications
include the use of a Mott state of bosonic atoms in an optical lattice
as a starting point to create controlled multiparticle entanglement as
an essential ingredient for quantum computation
\cite{Jaksch02,Mandel01,Briegel01}

The Mott-insulator quantum phase transition is driven by the interplay
of the repulsive interaction of bosons on the same lattice site and
the kinetic energy. Hence the ratio of the onsite energy and the
bandwidth forms the key parameter in the system. In optical lattices,
this parameter can be easily controlled and varied by several orders
of magnitude, enabling detailed studies of the quantum phase
transition. Probing the system by taking absorption pictures to image
the expansion patterns after a reasonable expansion time yields
information about the momentum distribution of the state. This
procedure was used to experimentally confirm the Mott transition
in an optical lattice \cite{Greiner01}.

The essential physics of cold bosonic atoms in an optical lattice is
captured by a bosonic Mott-Hubbard model describing the
competition between hopping and on-site interaction.  A number of
approximation schemes have been used to study this model analytically
\cite{Fisher01,Freericks01,Oosten01,Rey01} as well as numerically,
using approaches like 
the Gutzwiller mean-field ansatz \cite{Rokhsar01,Jaksch01},
density-matrix renormalization group (DMRG)
\cite{Kuehner01,Rapsch01,Kollath01}, exact diagonalization
(ED)\cite{Roth01,Roth02} and Quantum Monte Carlo (QMC)
\cite{Scalettar01,Batrouni01,Krauth,Kisker01,Kashurnikov01}. 

In this article, we study the short-range correlations, not
included by the Gutzwiller ansatz, by using perturbation theory. The
main purpose is to find corrections to the short-range behavior of the
one-particle density matrix, which is directly relevant to
experimentally observed expansion patterns.  These patterns are
important for determining the location of the insulator-superfluid
transition.  We note that in the insulating state our perturbative
approach is identical to the one used in \cite{Freericks01} (see also
\cite{Elstner}), although there the goal was different, viz., studying
corrections to the phase diagram.

The remainder of the article is organized as follows: In Section
\ref{modsec}, we will introduce the model and its mean-field
solution. The general perturbative approach is briefly outlined in
Section \ref{secPA}, while details may be found in the
Appendix. Numerical results are presented and discussed in Section
\ref{secNR}, first for local observables (\ref{secSLSO}) and then for
the density matrix (\ref{secRho}). Implications for expansion
patterns both for bulk systems and a harmonic confining potential
are discussed in Section \ref{harmotrap}.

\section{The model}\label{modsec}
The cold bosonic gas in the optical lattice can be described by a
Mott-Hubbard model \cite{Jaksch01}
\begin{equation}
H=\sum_{i=1}^{M}\frac{U}{2} n_i (n_i-1)
+\sum_{i=1}^{M}\left(\epsilon_i-\mu\right) n_i
-J\sum_{\langle i,j\rangle}  a_i^\dagger a_j
\label{MottHubb} \,.
\end{equation}
Here, $M$ is the total number of lattice sites, $a_i^\dagger$ ($a_i$)
creates (annihilates) a boson on site $i$, $n_i=a_i^\dagger a_i$, $U$
is the on-site repulsion describing the interaction between bosons on
the same lattice site, and $\mu$ denotes the chemical potential.  The
kinetic term includes only hopping between nearest-neighbor sites,
this is denoted by the summation index $\langle i,j \rangle$; $J$
is the hopping matrix element that we will assume to be lattice-site
independent. Finally, $\epsilon_i$ describes an
external on-site potential that is commonly present in experiments.

The Gutzwiller (GW) approach is based on an ansatz for the many-body
ground state that factorizes into single lattice-site wavefunctions
\begin{equation}
|G_0\rangle = \prod_{i=1}^{M}\left(\sum_{n=0}^{\infty} f_n^{(i)}
|n_i\rangle\right)\,.
\label{gutzwiller}
\end{equation}

The Gutzwiller wavefunction represents the ground state of the
following mean-field version of the Mott-Hubbard Hamiltonian,
Eq.~(\ref{MottHubb}):
\begin{equation}
\begin{split}
H_{\mathrm{MF}}&=\sum_{i=1}^{M}\frac{U}{2} n_i (n_i-1) 
+\sum_{i=1}^{M}\left(\epsilon_i-\mu \right) n_i\\
&- J\sum_{\langle
  i,j\rangle} \left(a_i^\dagger\Psi_j + \Psi_i^* a_j - \Psi_i^*
\Psi_j\right)\, . \label{MF}
\end{split}
\end{equation}
Here $\Psi_i$ is the mean-field potential on the $i$-th lattice site,
which is self-consistently defined as the expectation value of $a_i$
in terms of the Gutzwiller wavefunction, $\Psi_i=\langle
G_0|a_i|G_0\rangle$ \cite{Sheshadri01}.

Using the Gutzwiller ansatz to obtain an approximate variational
solution for the Mott-Hubbard Hamiltonian (\ref{MottHubb})
corresponds, however, to restricting the Hilbert space to the subset
of product states.  Consequently, even in higher dimensions, this
ansatz fails to describe the correct behavior of short-range
correlations between different lattice sites, which are important for
experimentally measurable observables, such as expansion patterns
(momentum distributions). Nevertheless, in the thermodynamic limit and
higher dimensions, the Gutzwiller wavefunction provides a good
approximation in the limits of $J\rightarrow 0$ and $U\rightarrow 0$
(i.e., deep in the Mott insulator (MI) and superfluid (SF) phases).
To get a satisfactory description of the short-range correlations we
will now derive perturbative corrections to the Gutzwiller mean-field
result.


\section{Perturbative Approach\label{secPA}}

Our aim is to start from the Gutzwiller approximation and improve it by
perturbatively including the short-range correlations between lattice sites.
We re-express the Mott-Hubbard Hamiltonian 
(\ref{MottHubb}) by adding the appropriate perturbation to the 
mean-field Hamiltonian, Eq.~(\ref{MF}):
\begin{equation}
H= H_{\mathrm{MF}}+V \, ,
\label{Hpert}
\end{equation}
with 
\begin{equation}
V= -J\sum_{\langle i,j\rangle}
 (a_i^\dagger-\Psi_i^*)(a_j-\Psi_j)\,.
\end{equation}
As the mean-field Hamiltonian represents a sum of single
lattice-site Hamiltonians, the excited states $|i_{\alpha}\rangle$ and
the excitation spectrum $\epsilon^{(i,\alpha)}$ can be obtained
numerically for each lattice site $i$ separately. Hence we can write
the excitations of $H_{\mathrm{MF}}$ as product states of 
single lattice-site excitations, 
\begin{equation}
|G^{\alpha}\rangle = \prod_{i=1}^{M} |i_{\alpha_i}\rangle \,, \label{ES}
\end{equation}
and the excitation spectrum as a sum over the 
single lattice-site excitation energies, 
\begin{equation}
\epsilon^{\alpha} = \sum_{i=1}^{M} \epsilon^{(i,\alpha_i)} \, , \label{ESp}
\end{equation}
where $\alpha=(\alpha_1,\alpha_2,\ldots)$ describes the set of
quantum numbers characterizing the given many-body energy eigenstate.

Having obtained the mean-field solution from the Gutzwiller ansatz, we
can now proceed to improve our wavefunction performing
Rayleigh-Schr\"odinger perturbation theory \cite{tannoudji01} in $V$:
\begin{equation}
\begin{split}
&\langle G^{\alpha} |G_l\rangle =\label{stateCorr}\\
&\frac{ \langle G^{\alpha} |V| G_{l-1}\rangle - 
\sum_{n=1}^{l-1}\epsilon_n\langle
 G^{\alpha}| 
G_{l-n}\rangle}{\epsilon_0 - \epsilon^{\alpha}}\,,
\end{split}
\end{equation}
where $|G_l\rangle$ and $\epsilon_l$ are the $l$-th order corrections to
the wavefunction and grand-canonical energy $E-\mu N$, respectively.

Knowing the excited wavefunctions, Eq.~(\ref{ES}), and the excitation
spectrum, Eq.~(\ref{ESp}), perturbative corrections to observables can
be calculated explicitly.  The resulting expressions up to second
order in $V$, which we use in the following, are derived and
illustrated diagrammatically in Appendix \ref{appendixA}.

\section{Numerical Results\label{secNR}}
We start by computing the Gutzwiller wavefunction numerically using a
conjugate-gradient descent method.  Propagation steps of the
time-dependent Gutzwiller equations \cite{Jaksch01} in imaginary time
are performed to test whether the minimum found by the conjugate-gradient
descent method is indeed the ground state.

Afterwards, we calculate eigenenergies and -vectors of each site, with
respect to the mean-field Hamiltonian. As explained above, this forms
the basis for perturbation theory, which we use to calculate the
corrections to the observables up to second order.  The results
obtained from perturbation theory show important modifications to
single-site observables as well as to the correlation function.

\subsection{Single Lattice-Site Observables\label{secSLSO}}        

These observables are composed of operators acting only on single
lattice sites. Hence, they describe local properties and 
will be less sensitive to the correlations between lattice
sites. Thus, values for single-site observables obtained from the
Gutzwiller wavefunction already provide a good approximation in most
cases (unless fluctuations are concerned).
\begin{figure}
\begin{center}
\includegraphics[width=8cm]{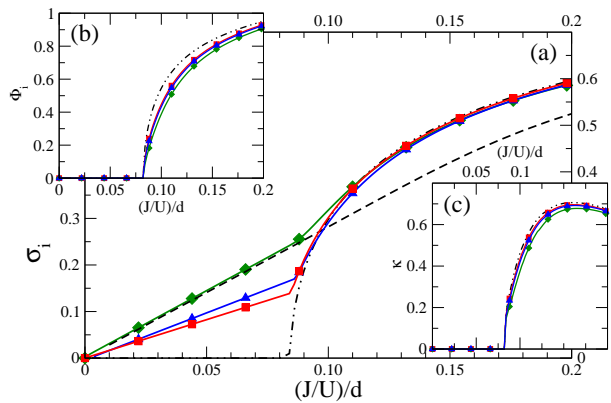}
\end{center}
\caption{Results from perturbation theory for homogeneous 
 lattices in $d=1$ (diamonds), $d=2$
 (triangles), and $d=3$ (squares) dimensions. The dashed-dotted lines
 in Fig.~\ref{slso}a-c are the Gutzwiller results.  (a) Number
 fluctuations $\sigma_i=\sqrt{\langle n_i^2\rangle - \langle
 n_i\rangle^2}$ calculated for a commensurate filling of one boson per
 lattice site. The dashed line shows the result from the exact
 diagonalization for 7 lattice sites and $N=7$ bosons. (b) Order
 parameter $\Phi_i=\langle a_i \rangle$ and (c) compressibility $\kappa$
 both computed at a fixed chemical potential $\mu /U=0.5$.
 \label{slso}}
\end{figure}

The leading corrections to the mean values of single lattice-site
observables (SLSO) are of second order in $V$.  The results for the
order parameter $\Phi_i$, the compressibility $\kappa$ and the number
fluctuations $\sigma_i$ are shown in Fig.~\ref{slso}, where the
ration $J/U$ has been scaled by the dimension, to keep the same
mean-field transition point. The solid lines in Fig.~\ref{slso} show
the results from perturbation theory, as compared to the Gutzwiller
result (shown as the dashed~-~dotted line). All three quantities show
a vanishing perturbative correction both for small and large $J/U$, as
the Gutzwiller wavefunction becomes a good approximation in these
regimes (for lattice dimensions $d>1$). As expected, deviations from
the mean-field picture are strongest near the MI-SF transition, where
higher-order corrections will become more and more important.

The order parameter $\Phi_i=\langle a_i\rangle$ shown in
Fig.~\ref{slso}b gets suppressed in the SF. Perturbative corrections
to the Gutzwiller result are particularly large in 1D (where the order
parameter vanishes in reality), but get smaller with increasing
dimension. This is not surprising, as Gutzwiller is a mean-field
approach and hence a better approximation for higher-dimensional
systems.  The critical value $(J/U)_c$ is not modified within the
present perturbative approach.

Figure~\ref{slso}c shows the results for the compressibility

\begin{equation}
\kappa = \frac{M}{N^2}\sum_{i=1}^M \frac{\partial \langle n_i\rangle}{\partial
  \mu} \, .
\end{equation}
The results of perturbation theory (PT) show a decrease of the
compressibility, pointing to an increasing stiffness of the SF phase
induced by the short-range interaction.

Finally, we computed the local particle number fluctuations $\sigma_i =
\sqrt{\langle n_i^2\rangle - \langle n_i\rangle^2}$. Exact
diagonalization calculations have shown that the number fluctuations
are changing smoothly at the MI-SF transition \cite{Roth01}, whereas
the Gutzwiller result predicts vanishing fluctuations in the MI phase,
$\sigma_i =0$ (see dashed-dotted line in Fig.~\ref{slso}c). Our
perturbative results reproduce the non-vanishing part of $\sigma_i$ in
the MI regime and agree well with our result obtained from exact
diagonalization of a one-dimensional system of $7$ lattice sites.
Significant deviations from the exact 1D result are seen in the MI-SF
transition regions, starting from the mean-field critical value,
$(J/U)_c=0.086$, up to values of the order of the critical values of
$(J/U)_c\approx 0.2\cdots 0.3$ usually obtained from DMRG
\cite{Rapsch01} and QMC \cite{Batrouni01} calculations.

Nevertheless, based on the good agreement of our perturbative result
with the exact diagonalization for a large region in the MI, we
conclude that the number fluctuations are mainly produced by
next-neighbor particle-hole fluctuations included in perturbation
theory.

\subsection{The Correlation Function\label{secRho}}

The single-particle density matrix $\rho_{ji}=\langle a_i^\dagger
a_j\rangle$  is of particular importance as it describes the
correlation between the different lattice sites. The correlation
function $\rho_{ij}$ shows off-diagonal long-range order in the SF
state (in dimensions $d>1$), in contrast to the MI phase where $\rho_{ij}$
 decays exponentially.

The experimental observation \cite{Greiner01} of the MI transition
relies on the different behavior of the density matrix in the MI and
SF regimes, which can be visualized by taking absorption pictures of
the freely expanding atomic cloud.  Assuming that the expansion time
is long enough and that the gas is dilute enough (such that atom-atom
interactions can be neglected during the expansion), the shape of the
cloud reflects the initial momentum distribution $\rho({\bf k})$,
which is directly given by the Fourier-transform of the density matrix
$\rho_{ij}$ :
\begin{equation}
\rho({\bf k})=|w({\bf k})|^2\sum_{i,j=1}^M \rho_{ij}e^{i{\bf k}({\bf r}_i
  -{\bf r}_j)}\, .
\label{fourier}
\end{equation}

\begin{figure}
\includegraphics[width=8cm]{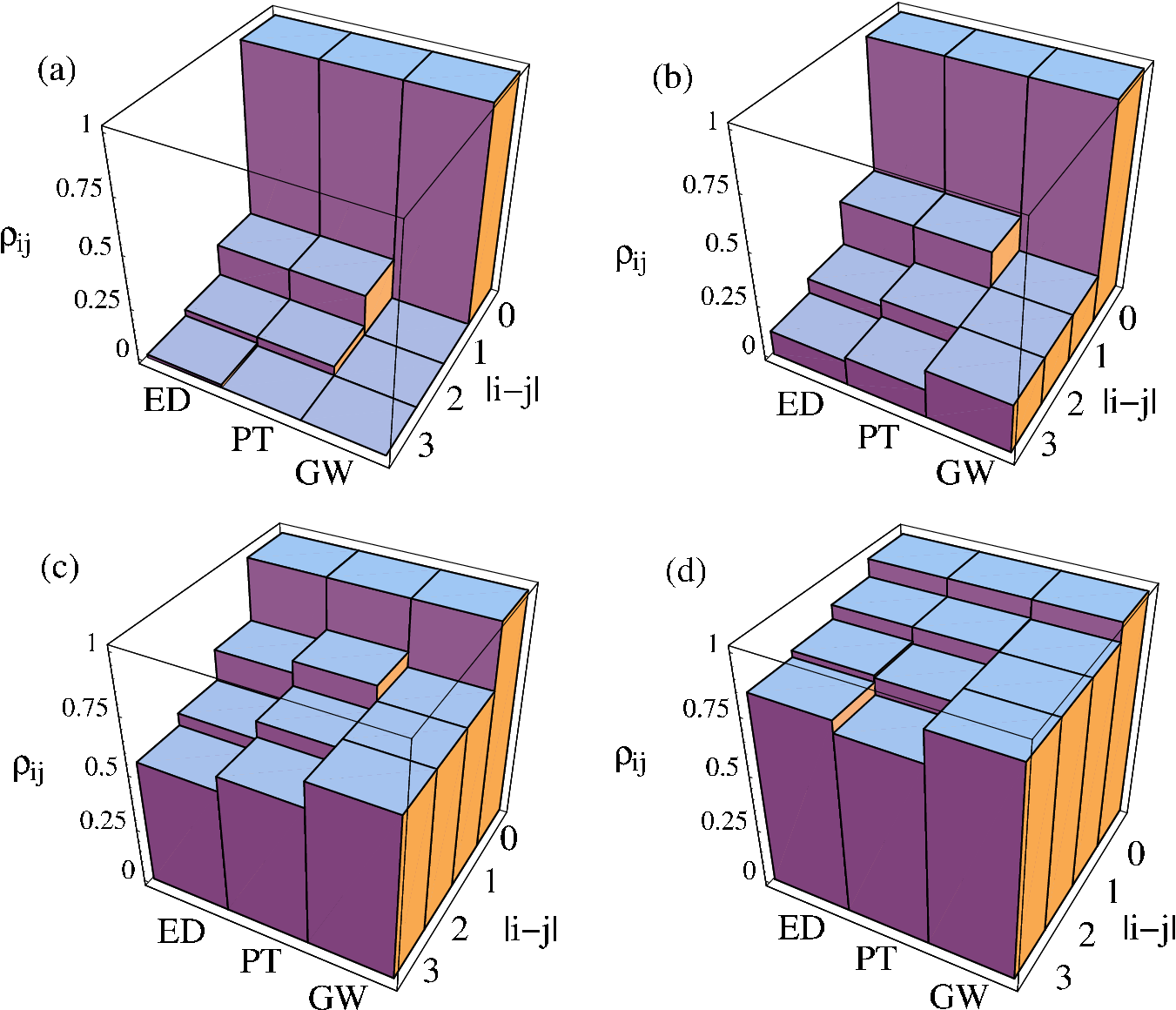}
\caption{Short-range behavior of the
  one-particle density matrix $\rho_{ij}$ as a function of site
  distance $|i-j|$. 
  Results for a homogeneous lattice of 7
  sites with $N=7$ Bosons have been obtained from
  exact diagonalization (ED), from
  second-order perturbation theory (PT), and from the Gutzwiller mean-field
  ansatz (GW).  (a) $J/U=0.05$ (MI regime), 
  (b) $J/U=0.1$, (c) $J/U=0.2$, (d) $J/U=0.5$ (deep SF regime).
 All calculations use periodic boundary conditions. 
  The mean-field value $(J/U)_c$ for the MI transition in the
  commensurate case with one boson per lattice site is
  $(J/U)_c=0.086$ in 1D (see Fig.~\ref{slso}).  (The MF value
  differs strongly from $(J/U)_c=0.277$ derived from QMC
  calculations \cite{Kuehner01} or $(J/U)_c=0.260$ for DMRG
  calculations \cite{Rapsch01}). 
\label{vglED}}
\end{figure}

Here, $w({\bf k})$ is the Fourier transform of the Wannier functions
$w({\bf r-r}_i)$ describing the wavefunction of a single lattice
site. The presence of the factor $w({\bf k})$ in Eq.~(\ref{fourier}) 
provides a cutoff at high momenta.

The mean-field results for $\rho_{ij}$ only describe the different
long-range behavior in the MI and SF. For a homogeneous lattice the
correlation function calculated from the Gutzwiller wavefunction gives
$\rho_{ii}=n_i$ for the diagonal elements and then drops instantly to
$\rho_{ij}= |\Phi|^2$ for all off-diagonal elements $i\neq j$. 
Short-range correlations are not reproduced by the Gutzwiller approach. This
deviation is particularly severe in the MI, where the
mean-field result predicts a completely flat momentum distribution,
whereas the short-range correlations (i.e. the exponential decay of
$\rho_{ij}$) yield smooth bumps in the
expansion pattern. These can be distinguished from the $\delta$-peaks
of the SF only after a sufficiently long expansion time.

Applying perturbation theory to the GW wavefunction improves the
structureless GW correlation function. In Fig.~\ref{vglED}, we have
compared the results of GW mean-field theory, of PT, and of exact
diagonalization (ED). The diagonalization has been carried out for a
small 1D lattice, where it is easily feasible. Although there is no
long-range order in the SF phase for the 1D case, where the density
matrix exhibits a power-law decay towards zero, it is still reasonable
to compare the short-range correlations. Indeed, we find a nice
agreement between perturbation theory and exact results, not only for
the MI (see Fig.~\ref{vglED}a), but also for the short-range behavior
in the SF.  This agreement is made possible by the fact that
$\rho_{ij}$ decays only slowly and higher-order corrections show only
negligible corrections for small lattices.  An example for the SF case
is shown in Fig.~\ref{vglED}d. However, there is still a considerable
difference for $\rho_{i,i+3}$ in Fig.~\ref{vglED}d. This is not
surprising as we do PT up to second order. Hence correlations over a
distance of three and more lattice sites are only corrected by the
global mean-field correction for the infinite lattice (see
Eq.~(\ref{CPsiOne}), Eq.~(\ref{CPsiTwo}), and Fig.~\ref{2ndCPsi} in
Appendix \ref{appendixA}). We expect better agreement for larger
lattice sizes as finite-size effects, arising in small lattices, are
still considerable for $M=7$ sites used in our ED calculations.

Finally, for intermediate values of $J/U$, shown in
Fig.~\ref{vglED}b,c, we observe a faster drop in the off-diagonal
correlations, such that higher-order contributions in the PT become
more important. In any dimension, the perturbation $V$ is no longer
small at the tip of the MI-SF transition lobe (Fig.~\ref{vglED}c), and
the PT breaks down.  Nevertheless, comparing with exact
diagonalization results, Fig.~\ref{vglED}b shows still good agreement
with ED, in contrast to Fig.~\ref{vglED}c, which shows clear
disagreement. Even though the parameters $J/U=0.2$ chosen for
Fig.~\ref{vglED}c are close to the MI-SF transition for 1D-lattices
(as predicted by DMRG \cite{Rapsch01} and QMC \cite{Kuehner01}
calculations), PT reproduces the correct slope for the off-diagonal
decay and lacks only the wrong offset from the mean field. Thus, even
for this case, PT represents a qualitative improvement on the
Gutzwiller result.  The results for both approximations, Gutzwiller
and PT, are expected to become better in higher dimensions (with the
perturbative corrections diminishing in size).

\begin{figure}
\begin{center}
\includegraphics[width=8cm]{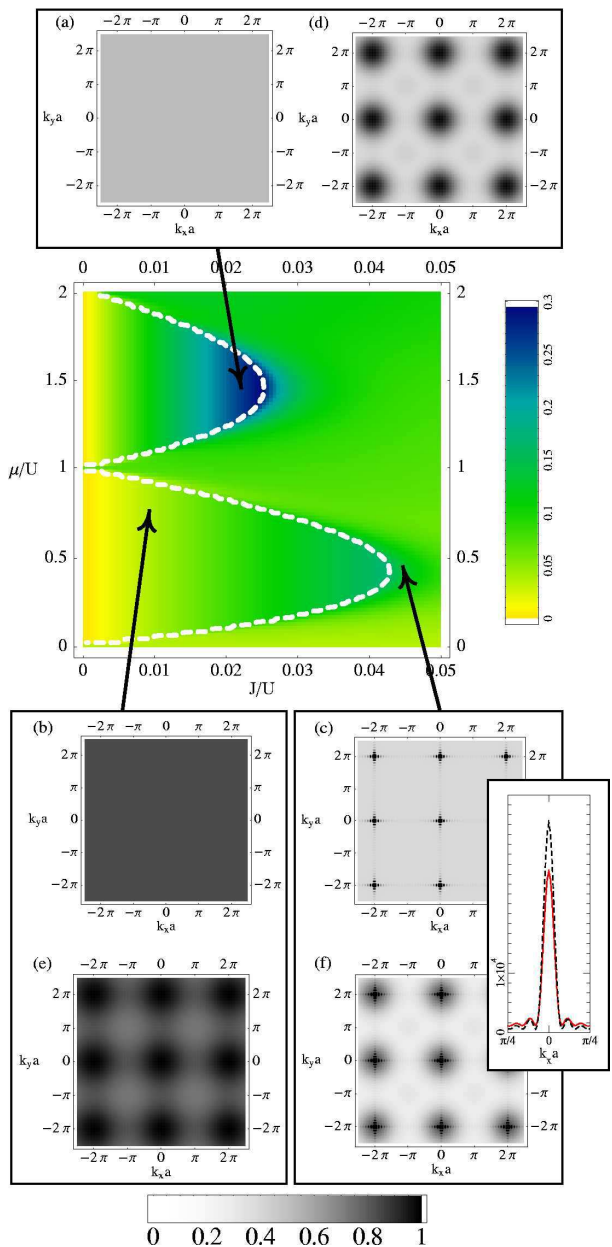}
\caption{The central figure shows the correction to $\rho_{i,i+1}$ in
  second-order PT as a function of $\mu/U$ and $J/U$. The order
  parameter $\langle \phi \rangle$ vanishes inside the Mott-insulating
  lobes, whose mean-field boundaries are given by the white dashed
  line. Plots (a) -- (f) display the resulting momentum distribution
  without the Wannier form factor, $\rho({\bf k})/|w({\bf k})|^2$,
  calculated for a 2D lattice with $25\times25$ lattice sites.
  (a)-(c) are the Gutzwiller mean-field results, and
  (d)-(f) are calculated using second-order PT.
  The inset in (c),(f) shows a
  cut of one peak taken along $k_ya=0$; dashed line for GW and 
  solid line for the   PT result.
  Arrows indicate the position of the respective plots in the
  $(\mu/U,J/U)$ phase diagram. The parameters used are: 
   $\mu/U=1.5$, $J/U=0.0225$
  for (a) and (d); $\mu/U=0.75$, $J/U=0.01$ for (b) and (e); and
  $\mu/U=0.5$, $J/U=0.044$ for (c) and (f). The gray-scales of plots
  belonging to the same parameter set are identical. Expansion patterns
  (a),(b),(d),and (e) are normalized to the peak maximum; (c) and (f) are
  normalized to $1/20$ of the peak maximum.\label{expans}
}
\end{center}
\end{figure}      

As discussed before, the perturbative enhancement of the description
of short-range correlations is expected to lead to strong consequences
for the momentum distributions. As an example we discuss a set of
momentum distributions $\rho({\bf k})/|w({\bf k})|^2$ for a
homogeneous 2D lattice, Fig.~\ref{expans}a-f, and compare the
Gutzwiller results to those improved by PT.  The improved PT versions,
Figs.~\ref{expans}d-f, show much finer structures than the mean-field
results, Fig.~\ref{expans}a-c. PT predicts broad peaks in the MI
regions down to very small values of $J/U$, Fig.~\ref{expans}e,
whereas the Gutzwiller result without PT shows a structureless flat
distribution for the whole MI region, Fig.~\ref{expans}a,b. Naturally,
the modifications of $\rho({\bf k})/|w({\bf k})|^2$ are strongest near
the phase transition, Figs.~\ref{expans}a and \ref{expans}d.

Going towards larger values $J/U$ into the SF phase, PT gives rise to
a suppression of the peaks (inset in Fig.~\ref{expans}c,f). This
suppression can be larger than $20\%$ of the original peak height and
stems from the corrections to the mean field.  Additionally,
Fig.~\ref{expans}f shows broad peaks induced by the inclusion of
short-range correlations. However, for large lattices, these broad
peaks are small compared to the (finite-size broadened) SF
$\delta$-peaks.


\subsection{Harmonic Trap Potential}
\label{harmotrap}
\begin{figure}
\begin{center}
\includegraphics[width=8cm]{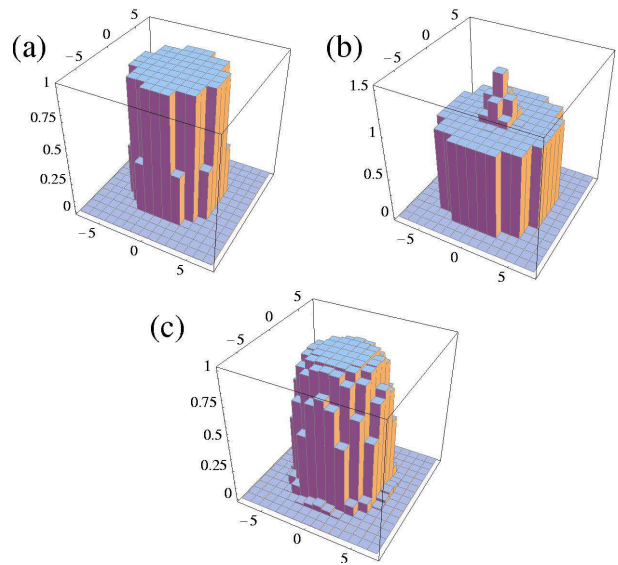}
\end{center}
\caption{Occupation number $n_i$ on a plane through the trap center for a
  3D lattice with $15^3$ lattice sites. (a) $\mu/U=0.4$, $J/U=0.005$, and
  $\alpha/U=0.02$. (b) $\mu/U=1.5$, $J/U=0.0075$, and
  $\alpha/U=0.03$. (c) $\mu/U=0.4$, $J/U=0.0075$, and
  $\alpha/U=0.02$.\label{spd}}
\end{figure}

\begin{figure}
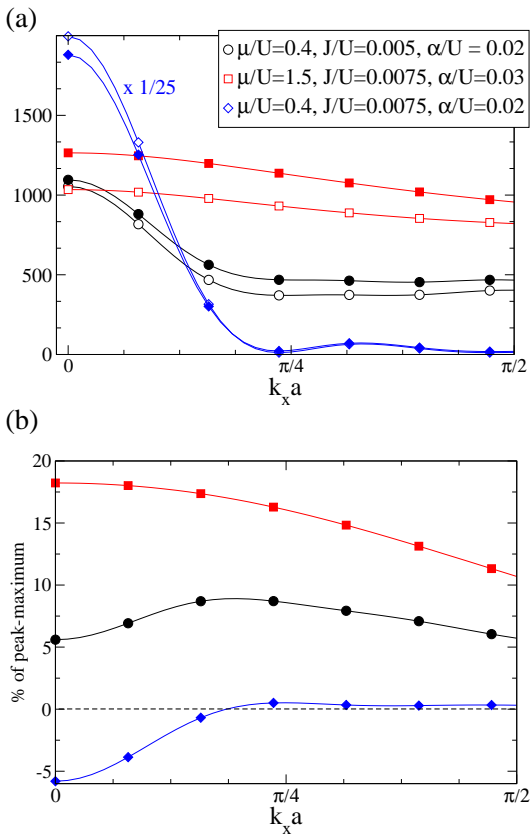

\begin{minipage}{8cm}
\includegraphics[width=7cm]{figure05.eps}
\includegraphics[width=7cm]{figure06.eps}
\end{minipage}
\caption{Expansion pattern for a 3D-lattice with $15^3$-sites in the
  presence of a harmonic potential. (a) Momentum distribution 
  without the Wannier form factor, $\rho({\bf k})/|w({\bf k})|^2$, calculated
  along the $k_x$ direction. Filled symbols are the PT results and open
  symbols are the GW results. The graphs for the MI surrounded by a SF shell
  (diamonds) are rescaled by a factor $1/25$. The results denoted by circles
  correspond to the occupation distribution of 
  Fig.~\ref{spd}a, diamonds to Fig.~\ref{spd}c, and squares to
  Fig.~\ref{spd}b.
(b) Difference between PT and GW results.
 \label{expans3D}}
\end{figure}

In contrast to what was assumed in the last section, optical lattices
used in experiments are not homogeneous. Magnetic or optical trapping
potentials are used \cite{Pedri01,Greiner01} to confine the atomic gas
to a finite volume.

The inhomogeneity caused by the trapping potential leads to slowly
varying on-site energies, $\epsilon_i$, in the Mott-Hubbard model
Eq.~(\ref{MottHubb}), that can be interpreted as a
spatially varying chemical potential $\mu_{\mathrm{local}}=\mu-\epsilon_i$. 
Consequently the lattice is in general not in a pure  MI or SF phase, but
shows alternating shells of SF and MI regions. An example for a SF region
surrounded by a MI shell is shown in Fig.~\ref{spd}b.

Considering the slowly varying on-site energy as a spatially varying
chemical potential gives a qualitative understanding of the shell
structures. The spatial variation of the chemical potential
corresponds to a path parallel to the $\mu/U$ axis in the
$\mu/U$-$J/U$-diagram. Starting with the potential minimum, in the
trap center, and then moving off the center, decreases the effective
local chemical potential. Whenever a
MI-SF (SF-MI) phase boundary is hit along the path in the
$\mu/U$-$J/U$-diagram, a change from a MI to a SF (SF to MI) shell
appears.

The inclusion of short-range correlations gives rise to considerable
modifications also in the presence of an inhomogeneous trapping
potential. Examples, calculated for a 3D-lattice, are shown in
Fig.~\ref{expans3D}, where an underlying harmonic potential
\begin{equation}
\epsilon_i = \alpha \sum_{\beta=1}^{3}i_\beta^2\,,
\end{equation}
was chosen.

The three situations considered here correspond to a case with a large
MI fraction (Fig.~\ref{spd}a), a SF island surrounded by a MI shell
(Fig.~\ref{spd}b), and a MI island surrounded by a SF phase
(Fig.~\ref{spd}c).  Calculating the expansion patterns for these
situations, Fig.~\ref{expans3D}, shows that the perturbative
corrections arising from the short-range correlations lead to
substantially different behavior in the different cases.

For the almost complete MI state we get a correction to all wavevectors
$k$, with a fast drop at values close to the peak center $k=0$, which
leads to a peak broadening in the expansion picture (see circles
in Fig.~\ref{expans3D}).

Particularly large changes were found for the case of a SF island
surrounded by a MI phase, Fig.~\ref{expans3D} (squares). Again,
corrections arise for all wavevectors, but, in contrast to the almost
homogeneous case, the largest increase is now found for $k=0$, with
changes of about $20\%$ of the peak-maximum.

Finally, the reversed situation, a MI island
surrounded by a SF phase (diamonds in Fig.~\ref{expans3D}) does
not show an increase of its maximum peak height but a considerable
reduction (over 5\% of the peak maximum). This is not surprising, as 
the majority of the lattice sites are now contributing to the SF phase,
and a peak reduction was also observed for the bulk SF phase.

\begin{figure}
\begin{center}
\includegraphics[width=8.5cm]{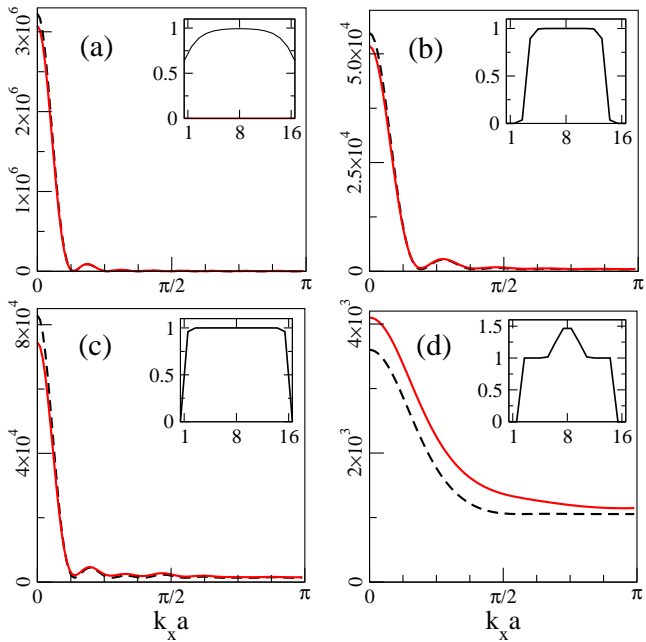}
\end{center}
\caption{Gutzwiller (dashed line) and perturbation theory (solid line)
  expansion patterns (cuts along $k_x$) for a system of $16^3$ lattice sites
  considered in \cite{Kashurnikov01}. All
 plots are without the Wannier form factor, $\rho({\bf{k}})/|w({\bf k})|^2$.
The insets show the occupation number $n_i$ for a cut along the
  $\hat{x}$-direction. (a)~$J/U = 1/32$, $\mu=0.3775$, $\alpha=0.00610$.
(b)~$J/U = 1/80$, $\mu=0.3125$, $\alpha=0.01221$. (c)~$J/U = 1/80$,
  $\mu=0.625$, 
  $\alpha=0.01288$. (d)~$J/U = 1/80$, $\mu=1$, $\alpha=0.02505$. (Note
  however, 
  that the Hamiltonian used in \cite{Kashurnikov01} differs from 
Eq.~(\ref{MottHubb})
  and hence the parameters are converted to the  corresponding quantities used
  in our definitions.) \label{Kash}}
\end{figure}

We would now like to compare our approach with the results obtained by
Kashurnikov {\it et al.} \cite{Kashurnikov01}, who used QMC
calculations to calculate expansion patterns for a small 3D lattice
with harmonic confinement. There is good qualitative agreement in all
cases with high superfluid fraction (compare for example 
Fig.~\ref{Kash}a(b) and Fig.~2b(c) in Ref. \cite{Kashurnikov01}).
 However, the features of these expansion
patterns are already well reproduced using the Gutzwiller mean-field
ansatz alone, in particular, the satellite peak, which was discussed
as a signature of the MI-SF shell structure in \cite{Kashurnikov01}.
  Corrections arising from PT show a
suppression of the SF peak, as was discussed above.

However, there are also considerable discrepancies to the
QMC results for situations with a large MI fraction, 
even after implementing second-order PT.  
In these cases (Fig.~\ref{Kash}c(d) and Fig.~2d(e) in Ref. \cite{Kashurnikov01}), the influence
of the MI phase on the expansion picture broadens the peak and leads
to a homogeneous background. The discrepancies to the QMC are clearly visible
in Fig.~\ref{Kash}c showing no peak broadening and a sattelite peak in
contrast to the QMC results (Fig.2d in \cite{Kashurnikov01}). Including the
short range correlations perturbatively corrects the expansion pattern in the
right direction, giving rise to a suppression of the SF-peak. 
Considering the expansion pattern with
the clearest MI features (Fig~\ref{Kash}d and Fig.~2e in
Ref. \cite{Kashurnikov01}), we obtain the correct peak broadening from GW/PT
calculation, but a larger ratio of MI background to SF peak.

Concerning these discrepancies, we note that the expansion pattern is
highly sensitive to the value of the mean field. Even small deviations
can lead to a change in the SF-peak height sufficient to mask the flat
distribution of the MI phase. We checked, however, that the observed
discrepancy is not due to a lack in accuracy of our numerical
calculations.  We therefore believe that the discrepancies between QMC
and GW/PT for situations with a large MI fraction can be attributed to
the insufficiency of GW and low-order PT in describing the long-range
correlations in this inhomogeneous situation.
In addition, we note that the lattice employed in \cite{Kashurnikov01}
is comparatively small for the given harmonic confinement potential
(with no complete shell of empty sites at the perimeter, see insets of 
Fig.~\ref{Kash}), i.e., the
choice of boundary conditions (periodic in the case of our numerical
calculations) may have non-negligible effects on the outer lattice
sites.

\section{Conclusions}

We have used perturbation theory to incorporate the effects of 
short-range correlations on top of the mean field solution of the bosonic
Mott Hubbard model.  We derived corrections to local quantities, as
well as to observable expansion patterns.  We numerically calculated
the corrections to the MF-result, using PT up to second order, thus
including correlations between next and next-nearest lattice sites.
Modifications to the particle number fluctuations $\sigma_i$, arising
from the PT, gave rise to the expected smooth transition of $\sigma_i$
at the MI-SF transition. Moreover, comparing the PT results to the ED
results for $\sigma_i$ in 1D lattices showed good agreement for small
values of $J/U$.  Of particular importance are the corrections to the
correlation function, and thus to the expansion patterns.  We compared
the correlation function obtained from PT with calculations from ED
for small 1D-lattices and found good agreement.  Studying the
expansion patterns showed that the inclusion of the short-range
correlations to the MF-ansatz gives rise to distinct modifications. A
broad peak can be seen in the PT results for the MI regime, visible
even down to small values of $J/U$. Comparing PT and MF expansion
patterns obtained for parameters in the SF region displayed a
considerable suppression of the SF peak in the PT
results. Additionally, on approaching the SF-MI transition from the SF
side, broad peaks underlying the SF peaks were found in the
PT-expansion patterns.  Including a harmonic confinement potential
leads to situations where SF and MI regions coexist. Hence the
perturbative corrections to the expansion pattern become more
complex. We studied lattices with different underlying harmonic traps
giving rise to different constellations of SF-MI regions, finding
modifications of up to $20\%$ of the peak maximum in the expansion
patterns.


\begin{acknowledgments} 
We would like to thank Krishnendu Sengupta for discussions.
Our work was supported by the Swiss NSF and the NCCR Nanoscience, as well as
a DFG research fellowship (F.M.).
\end{acknowledgments}

\appendix

\begin{figure}
\includegraphics[width=7cm]{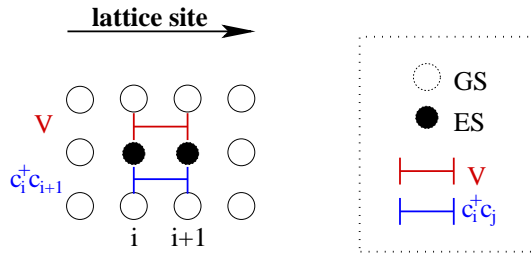}
\caption{\label{diagramm} Schematic diagram illustrating the term
  $\mv{c_i^\dagger c_j\frac{1}{\Delta}V_{ij}}$ appearing in the first-order
 correction ${\langle c_i^\dagger c_j \rangle}_1$ of the density matrix, where
  $V_{ij}=\frac{J}{2}(c_i^\dagger c_j  + c_j^\dagger c_i)$ is the 
  perturbation connecting   sites $i$ and $j$.  The
  diagram shows the lattice sites in the horizontal direction. The
  different steps needed to obtain the matrix element are shown
  vertically. Open circles denote the ground state (GS)
  of the given lattice site, while filled circles are excited 
  states (ES) of this
  site, for the mean-field Hamiltonian $H_{\mathrm{MF}}$.}
\end{figure}

\section{Derivation of the Matrix Elements \label{appendixA}}

We use standard stationary perturbation theory \cite{tannoudji01} to
calculate the corrections to the mean-field results induced by the
perturbation
\begin{equation}
V=-\sum_{\langle i,j\rangle} V_{ij} =-\frac{J}{2}\sum_{\langle i,j\rangle}
(c_i^\dagger c_j  + c_j^\dagger c_i)\,,\label{pert}
\end{equation}
where we introduce the new operators 
\begin{equation}
c_i = a_i -\langle a_i \rangle_0 = a_i -\Psi_i \,,
\end{equation}
The expectation value $\langle\cdot\rangle_0$ is
taken with respect to the MF
wavefunction $|G_0\rangle$. Defining the operator for the energy denominator
as
\begin{equation}
\frac{1}{\Delta}\equiv \frac{1-|G_0\rangle\langle G_0|}{\epsilon_0 -
H_{\mathrm{MF}}}\,,
\end{equation}  
the expectation value of an observable $\langle A\rangle$ including
all corrections up to second order is
\begin{equation}
\begin{split}
\langle A \rangle  \approx & \mv{A} + \mv{V\frac{1}{\Delta}A} +
\mv{A\frac{1}{\Delta}V}\\
& + \mv{V\frac{1}{\Delta}V\frac{1}{\Delta}A} +
\mv{V\frac{1}{\Delta}A\frac{1}{\Delta}V}\\ & +
\mv{A\frac{1}{\Delta}V\frac{1}{\Delta}V}
 - \mv{V\frac{1}{\Delta^2}V}\mv{A} \,. \label{pertA}
\end{split}
\end{equation}
The first line in Eq.~(\ref{pertA}) is the MF-result $\mv{A}$, followed by two
contributions which are the first-order corrections. Lines two and three in
Eq.~(\ref{pertA}) are the second-order corrections to the mean value.

\subsection{Density Matrix: First-Order Corrections}

As an example we will discuss the corrections to the density matrix
$\rho_{ij}=\langle a_i^\dagger a_j\rangle$. The first-order correction to the
density matrix is: 
\begin{equation}
\langle a_i^\dagger a_j\rangle_1=\langle c_i^\dagger
c_j\rangle_1 + \Psi_i^*\langle c_j\rangle_1 + \langle
c_i^\dagger\rangle_1\Psi_j + \Psi_i^*\Psi_j \,. \label{AtoC}
\end{equation}
For two different lattice sites $i \neq j$, we find
\begin{equation}
\mv{c_i^\dagger c_j \frac{1}{\Delta} c_i} =
\mv{c_j}\mv{c_i^\dagger\frac{1}{\Delta} c_i} = 0 \, ,
\end{equation}
since the Gutzwiller ground state is a product state. As a consequence, 
the contributions  $\langle c_i^\dagger \rangle_1$ and
$\langle c_j \rangle_1$ vanish. The only remaining contributions to $\langle
a_i^\dagger a_j\rangle_1$ stem from
\begin{equation}
\langle c_i^\dagger c_j\rangle_1 = -\mv{V_{ij}\frac{1}{\Delta}c_i^\dagger c_j}
-\mv{c_i^\dagger c_j \frac{1}{\Delta}V_{ij}} \,, \label{firstorder}
\end{equation} 
and $\Psi_i^*\Psi_j$.

To get a better idea of the character of the terms arising in
the perturbative expansion, we introduce, in Fig.~\ref{diagramm}, a
graphical representation (for the example of a 1D-lattice). 

The graph shows a
decomposition of the matrix element, with each row showing the wavefunction
at an intermediate step in the evaluation of the matrix element. 
As we deal with  product states
\begin{equation}
|G^\alpha\rangle = \prod_{i=1}^M |i_{\alpha_i}\rangle\,,
\end{equation}
we represent the wavefunction by a row of circles, where each circle denotes the state $|i_{\alpha_i}\rangle$ of a particular
lattice site $i$.

Open circles
in Fig.~\ref{diagramm} denote a lattice site in its ground state (GS),
filled circles refer to an excited state (ES) of this particular
lattice site, with respect to the local mean-field Hamiltonian. Note
that, in general, this can be an arbitrarily highly excited state
(although higher contributions are suppressed by the energy
denominator, and a cutoff is used in practice).

Starting with a row of open circles, denoting the GS, $|G_0\rangle$,
each following row corresponds to the state after the action of $V$ or
$c_i^\dagger c_j$, as indicated on the left side of the graph.  As all
matrix elements in Eq.~(\ref{pertA}) can be expressed in terms of a GS
expectation value $\langle\cdot\rangle_0$ and a sequence of $V$ and
$c_i^\dagger c_j$ operators, the first and last row must always be a
line of open circles.

Let us consider for instance the second term in Eq.~(\ref{firstorder})
\begin{equation}
\mv{c_i^\dagger c_j\frac{1}{\Delta}V_{ij}}\,.
\end{equation} 
Reading the graph in Fig.~\ref{diagramm} from top to bottom corresponds to
reading the matrix element from right to left. Starting with the GS, $|
G_0\rangle$, the first row consists of open circles. The second row shows the
state after the action of the perturbation $V$. 
Acting with $V_{ij}$ to the
right onto the GS results in a state 
\begin{equation}
V_{ij} |G_0\rangle = \sum_{\alpha,\beta} f_{\alpha,\beta}
| i_\alpha,j_\beta \rangle \,,
\end{equation}
where $i$ and $j$ are neighboring lattice sites and $|i_\alpha ,j_\beta\rangle$
denotes the state with lattice site $i$ ($j$) 
in the exited state $\alpha$ ($\beta$) and all other sites in their GS. Thus
the second row shows the lattice sites $i$ and $i+1$ in an excited state
(filled circle),
as
the perturbation, Eq.~(\ref{pert}), allows only next neighbor interactions.
Finally, the action of $c_i^\dagger c_j$ has to bring the 
excited states back to the GS, in order to get a non-vanishing
contribution.
Therefore, in first order PT,  only next neighbor corrections
 to the
correlation function arise, as the final row  must represent
the ground state $\langle G_0|$ again.
The graph representing the remaining first term in Eq.~(\ref{firstorder}) is
obtained by rotating the graph in Fig.~\ref{diagramm} by $\pi$.

\begin{figure}
\includegraphics[width=8cm]{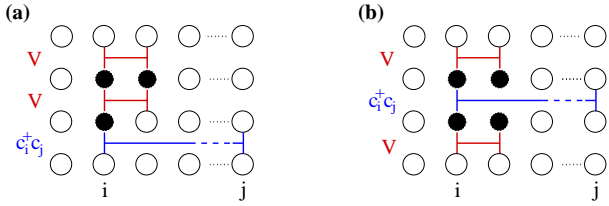}
\caption{Diagrammatic representation of the contributions from
  $\Psi_i^*\langle c_j\rangle_2$ and $\langle
  c_i^\dagger\rangle_2\Psi_j$ to the second-order correction of the 
  density matrix. (a)  Terms representing
  contributions of the form Eq.~(\ref{CPsiOne}). (b)
  contributions of the form Eq.~(\ref{CPsiTwo}). All
  notations are the same as in Fig.~\ref{diagramm}. \label{2ndCPsi}}
\end{figure}

\subsection{Density Matrix: Second-Order Corrections}

Rewriting the second-order corrections to the density matrix, $\langle
a_i^\dagger a_j\rangle_2$, in terms of the operators $c_i^\dagger$ and
$c_j$ gives Eq.~(\ref{AtoC}), but with $\langle\cdot\rangle_1$
replaced by $\langle\cdot\rangle_2$. In contrast to the first-order
corrections, now the terms proportional to $\langle c_j\rangle_2$ and
$\langle c_i^\dagger\rangle_2$ also give non-vanishing contributions.
Using Eq.~(\ref{pertA}) we obtain:
\begin{eqnarray}
\langle c_j\rangle_2&=
\sum_{k}^{\prime}\mv{V_{jk}\frac{1}{\Delta}V_{jk}\frac{1}{\Delta}c_j}
+ \sum_{k}^{\prime}\mv{c_j\frac{1}{\Delta}V_{jk}\frac{1}{\Delta}V_{jk}}& 
\qquad \label{CPsiOne}\\
 &+ \sum_{k}^{\prime}\mv{V_{jk}\frac{1}{\Delta}c_j\frac{1}{\Delta}V_{jk}} \,.&
\label{CPsiTwo}
\end{eqnarray}
Here, the primed sums run over all neighbors $k$ to site $j$.
The corresponding subset of graphs for $\Psi_i^*\langle c_j\rangle_2$ and
$\langle c_i^\dagger\rangle_2\Psi_j$ are given by Fig.~\ref{2ndCPsi}a and
Fig.~\ref{2ndCPsi}b for Eq.~(\ref{CPsiOne}) and Eq.~(\ref{CPsiTwo})
respectively. Note that all terms of Eq.~(\ref{CPsiOne}) and
Eq.~(\ref{CPsiTwo}) give a correction to all matrix elements of the density
matrix independent of the distance between the lattice sites. We can
understand these contributions as a modification to the MF value of
the density matrix. 

\begin{figure}
\includegraphics[width=7cm]{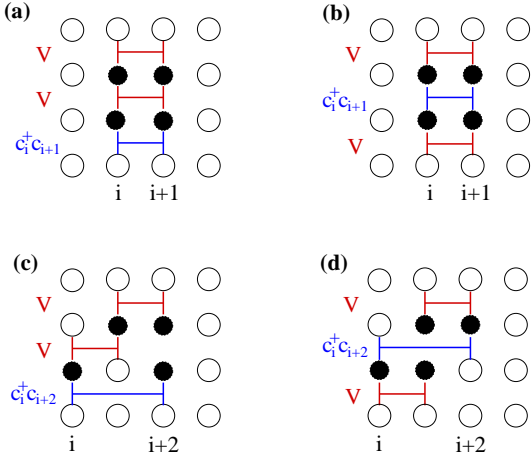}
\caption{\label{2ndCC} Graphs showing second-order corrections arising
  from $\langle c_i^\dagger c_j\rangle_2$. Diagrams (a) and (b) are
  coming from direct-neighbor contributions as given by
  Eq.~(\ref{2ndCCOne}) and Eq.~(\ref{2ndCCTwo}) respectively. Diagrams
  (c) and (d) are next-nearest-neighbor contributions: (c)
  corresponds to Eq.~(\ref{2ndCCThree}) and Eq.~(\ref{2ndCCFour}); 
  (d) corresponds to Eq.~(\ref{2ndCCFive}).}
\end{figure}

For the second-order contribution, $\langle c_i^\dagger c_j\rangle_2$, we have
to distinguish two cases:
\begin{itemize}
\item[(a)] Lattice site $i$ and $j$ being direct neighbors. In this case we
  get
\begin{eqnarray}
\langle c_i^\dagger c_j\rangle_2 =&
\mv{V_{ij}\frac{1}{\Delta}V_{ij}\frac{1}{\Delta}c_i^\dagger c_j} +
\mv{c_i^\dagger c_j \frac{1}{\Delta}V_{ij}\frac{1}{\Delta}V_{ij}}
&\qquad\label{2ndCCOne}\\ 
& + \mv{V_{ij}\frac{1}{\Delta} c_i^\dagger
c_j\frac{1}{\Delta}V_{ij}}\label{2ndCCTwo} \,. &
\end{eqnarray}
Corrections for Eq.~(\ref{2ndCCOne}) and Eq.~(\ref{2ndCCTwo}) are shown in
Fig.~\ref{2ndCC}a and Fig.~\ref{2ndCC}b, respectively.

\item[(b)] Configurations corresponding to two lattice sites 
$i \ne j$ connected by two successive hopping steps via site $k$. This
gives rise to six contributions:
\begin{small}
\begin{eqnarray}
&&\langle c_i^\dagger c_j\rangle_2 = \qquad\\
&&\sum_{\mathcal{C}}\mv{V_{jk}\frac{1}{\Delta}V_{ki}\frac{1}{\Delta}c_i^\dagger
c_j} +
\sum_{\mathcal{C}}\mv{V_{ki}\frac{1}{\Delta}V_{jk}\frac{1}{\Delta}c_i^\dagger
c_j}\qquad\label{2ndCCThree}\\ 
&&+ \sum_{\mathcal{C}}\mv{c_i^\dagger
c_j\frac{1}{\Delta}V_{jk}\frac{1}{\Delta}V_{ki}} +
\sum_{\mathcal{C}}\mv{c_i^\dagger
c_j\frac{1}{\Delta}V_{ki}\frac{1}{\Delta}V_{jk}}\qquad\label{2ndCCFour}\\
&&+ \sum_{\mathcal{C}}\mv{V_{jk}\frac{1}{\Delta}c_i^\dagger
c_j\frac{1}{\Delta}V_{ki}} +
\sum_{\mathcal{C}}\mv{V_{ki}\frac{1}{\Delta}c_i^\dagger
c_j\frac{1}{\Delta}V_{jk}}\,.\qquad\label{2ndCCFive}
\end{eqnarray}
\end{small}
An example for the contributions arising from Eq.~(\ref{2ndCCThree})
and Eq.~(\ref{2ndCCFour}) is shown in Fig.~\ref{2ndCC}c. The last
term, Eq.~(\ref{2ndCCFive}), has the representation shown in
Fig.~\ref{2ndCC}d.  Note that for lattices with dimensions
$d>1$, the sites $i$,$j$ and $k$ need not necessarily form a straight
line but can form a chevron.  
\end{itemize}



\begin{thebibliography}{99}

\bibitem{Fisher01} M.~P.~A. Fisher, P.~B. Weichman, G. Grinstein, and
  D.~S. Fisher, Phys. Rev. B {\bf 40}, 546 (1989).

\bibitem{Jaksch01} D. Jaksch, C. Bruder, J.~I. Cirac, C.~W. Gardiner,
 and P. Zoller, Phys. Rev. Lett. {\bf 81}, 3108 (1998).

\bibitem{Greiner01} M. Greiner, O. Mandel, T. Esslinger, T.~W. H\"ansch, and
  I. Bloch, Nature (London) {\bf 415}, 39 (2002).

\bibitem{Jaksch02} D. Jaksch, H.-J. Briegel, J.~I. Cirac, C.~W. Gardiner, and
  P. Zoller, Phys. Rev. Lett. {\bf 82}, 1975 (1999).

\bibitem{Jaksch03} D. Jaksch, V. Venturi, J.~I. Cirac, C.~J. Williams, 
and P. Zoller, 
Phys. Rev. Lett. {\bf 89}, 040402 (2002). 

\bibitem{Hofstetter} W. Hofstetter, J.~I. Cirac, P. Zoller, E. Demler,
  and M.~D. Lukin,
Phys. Rev. Lett. {\bf 89}, 220407 (2002).

\bibitem{Altman} E. Altman and A. Auerbach,
Phys. Rev. Lett. {\bf 89}, 250404 (2002). 

\bibitem{Zwerger01} W. Zwerger, J. Opt. B: Quantum
Semiclass. Opt. {\bf 5}, S9 (2003).

\bibitem{Buechler01} H.~P. B\"uchler, G. Blatter, and W. Zwerger,
Phys. Rev. Lett. {\bf 90}, 130401 (2003).

\bibitem{Burnett} D.~C. Roberts and K. Burnett,
Phys. Rev. Lett. {\bf 90}, 150401 (2003).

\bibitem{Kleinert01} H. Kleinert, S. Schmidt, and A. Pelster, cond-mat/0307412.

\bibitem{Damski} B. Damski, J. Zakrzewski, L. Santos, P. Zoller, 
and M. Lewenstein, 
Phys. Rev. Lett. {\bf 91}, 080403 (2003).  

\bibitem{Jaksch04} P. Rabl, A.~J. Daley, P.~O. Fedichev, J.~I. Cirac,
  and P. Zoller,
Phys. Rev. Lett. {\bf 91}, 110403 (2003).  

\bibitem{Buechler02} H.~P. B\"uchler and G. Blatter, 
Phys. Rev. Lett. {\bf 91}, 130404 (2003).

\bibitem{Lewenstein} M. Lewenstein, L. Santos, M.~A. Baranov, and H. Fehrmann, 
Phys. Rev. Lett. {\bf 92}, 050401 (2004)  

\bibitem{Greiner02} M. Greiner, O. Mandel, T.~W. H\"ansch, and
  I. Bloch, Nature (London) {\bf 419}, 51 (2002).

\bibitem{Mandel01} O. Mandel, M. Greiner, A. Widera, T. Rom, T.~W. H\"ansch,
  and I. Bloch, Nature (London) {\bf 425}, 937 (2003).

\bibitem{Esslinger} T. St\"oferle, H. Moritz, C. Schori, M. K\"ohl,
  and T. Esslinger,
Phys. Rev. Lett. {\bf 92}, 130403 (2004).

\bibitem{Briegel01} H.-J. Briegel and R. Raussendorf, Phys. Rev. Lett. {\bf
  86}, 910 (2001).

\bibitem{Freericks01} J.~K. Freericks and H. Monien, Phys. Rev. B {\bf
53}, 2691 (1996).

\bibitem{Oosten01} D. van Oosten, P. van der Straten, and H.~T.~C. Stoof,
  Phys. Rev. A {\bf 63}, 053601 (2001).

\bibitem{Rey01} A.~M. Rey, K. Burnett, R. Roth, M. Edwards, C.~J. Williams,
  and C.~W. Clark, J. Phys. B {\bf 36}, 825 (2003).

\bibitem{Rokhsar01} D.~S. Rokhsar and B.~G. Kotliar, Phys. Rev. B {\bf
 44}, 10328 (1991).

\bibitem{Kuehner01} T.~D. K\"uhner and H. Monien, Phys. Rev. B {\bf 58},
  R14741 (1998).

\bibitem{Rapsch01} S. Rapsch, U. Schollw\"ock, and W. Zwerger,
  Europhys. Lett. {\bf 46}, 559 (1999).

\bibitem{Kollath01} C. Kollath, U. Schollw\"ock, J. von Delft, and
  W. Zwerger, Phys. Rev. A {\bf 69}, 031601 (2004).

\bibitem{Roth01} R. Roth and K. Burnett, Phys. Rev. A {\bf 68}, 023604
  (2003).

\bibitem{Roth02} R. Roth and K. Burnett, Phys. Rev. A {\bf 67}, 031602(R)
  (2003).

\bibitem{Scalettar01} R.~T. Scalettar, G.~G. Batrouni, and G.~T. Zimanyi,
  Phys. Rev. Lett. {\bf 66}, 3144 (1991).

\bibitem{Batrouni01} G.~G. Batrouni and R.~T. Scalettar, Phys. Rev. B {\bf 46},
  9051 (1992). 

\bibitem{Krauth} W. Krauth and N. Trivedi, Europhys. Lett. {\bf 14},
  627 (1991); W. Krauth, N. Trivedi and D. Ceperley,
  Phys. Rev. Lett. {\bf 67}, 2307 (1991).

\bibitem{Kisker01} J. Kisker and H. Rieger, 
Phys. Rev. B {\bf 55}, R11981 (1996).

\bibitem{Kashurnikov01}  V.~A. Kashurnikov, N.~V. Prokof'ev, and 
  B.~V. Svistunov,  Phys. Rev. A {\bf 66}, 031601(R) (2002).

\bibitem{Elstner} N. Elstner and H. Monien, cond-mat/9905367 (unpublished).

\bibitem{Sheshadri01} K. Sheshadri, H.~R. Krishnamurthy, R. Pandit,
and T.~V. Ramakrishnan, Europhys. Lett. {\bf 22}, 257 (1993).


\bibitem{Pedri01} P. Pedri {\it et al.}, 
Phys. Rev. Lett. {\bf 87}, 220401 (2001).

\bibitem{tannoudji01} C. Cohen-Tannoudji, B. Diu, and F. Lalo\"e,
{\it Quantum Mechanics} (Hermann, Paris, 1977).

\end{thebibliography}
\end{document}